\documentclass[12pt]{article}

\pdfoutput=1

\usepackage{graphicx}
\usepackage{caption}
\usepackage{subcaption}
\usepackage{sidecap}
\usepackage{wrapfig}
\usepackage{amsmath}
\usepackage{amssymb}
\usepackage{cite}
\usepackage{hyperref}

\newcommand\pubnumber{}
\newcommand\pubdate{\today}

\def\institute{CP$^{3}$-Origins, University of Southern Denmark, Campusvej 55, DK-5230 Odense M, Denmark}
\def\support{\footnote{DNRF90}}

\def\Title#1{\begin{center} {\Large #1 } \end{center}}
\def\Author#1{\begin{center}{ \sc #1} \end{center}}
\def\Address#1{\begin{center}{ \it #1} \end{center}}

\newcommand\pubblock{\rightline{\begin{tabular}{l} \pubnumber\\
         \pubdate  \end{tabular}}}
\newenvironment{Abstract}{\begin{quotation}  }{\end{quotation}}
\newenvironment{Presented}{\begin{quotation} \begin{center} 
             PRESENTED AT\end{center}\bigskip 
      \begin{center}\begin{large}}{\end{large}\end{center} \end{quotation}}

\begin{document}
\begin{titlepage}
\pubblock

\vfill
\Title{Top Signatures From Composite Higgs Theories}
\vfill
\Author{ Natascia Vignaroli\support}
\Address{\institute}
\vfill
\begin{Abstract}
Many compelling theories to address the Higgs hierarchy problem predict strong interactions between the top and a sector of New Physics. In minimal composite Higgs models (CHM), the top interactions with a BSM strongly-interacting sector give the leading contribution to trigger the EWSB and generate a light mass for the pseudo-NG Higgs. This implies that new composite states (vector-like top-partners, composite vector resonances, new composite scalars) dominantly interact with third-generation quarks and, when produced at colliders, generate top quarks in the final state. I will indicate interesting signatures involving top quarks for CHM discovery/test at the LHC and future colliders.
\end{Abstract}
\vfill
\begin{Presented}
$9^{th}$ International Workshop on Top Quark Physics\\
Olomouc, Czech Republic,  September 19--23, 2016
\end{Presented}
\vfill
\end{titlepage}
\def\thefootnote{\fnsymbol{footnote}}
\setcounter{footnote}{0}

\section{Introduction}

Composite Higgs models are very compelling theories to address the Higgs hierarchy problem. The electroweak symmetry breaking (EWSB) is triggered by a new strong dynamics, composite at the TeV scale, from which the Higgs emerges as a composite state. The composite Higgs can be significantly lighter than other composite resonances (which have masses of $\mathcal{O}(1)$ TeV) if it is also the Nambu-Goldstone boson associated to the breaking of a global symmetry, $\mathcal{G}$, of the strong sector \cite{Kaplan:1983fs}. The Higgs potential is generated by an explicit breaking of $\mathcal{G}$. In minimal realizations of the composite Higgs paradigm \cite{Agashe:2004rs} the leading source of the breaking is provided by the top interactions with the strong sector. The implications are that we expect vectorlike quarks (VLQ) top-partners with masses $\lesssim$ 1 TeV \cite{Matsedonskyi:2012ym} and large couplings between the top and the new composite resonances.   

\section{Top-partner VLQs}

\begin{figure}[htb]
\centering
\includegraphics[height=1.3in]{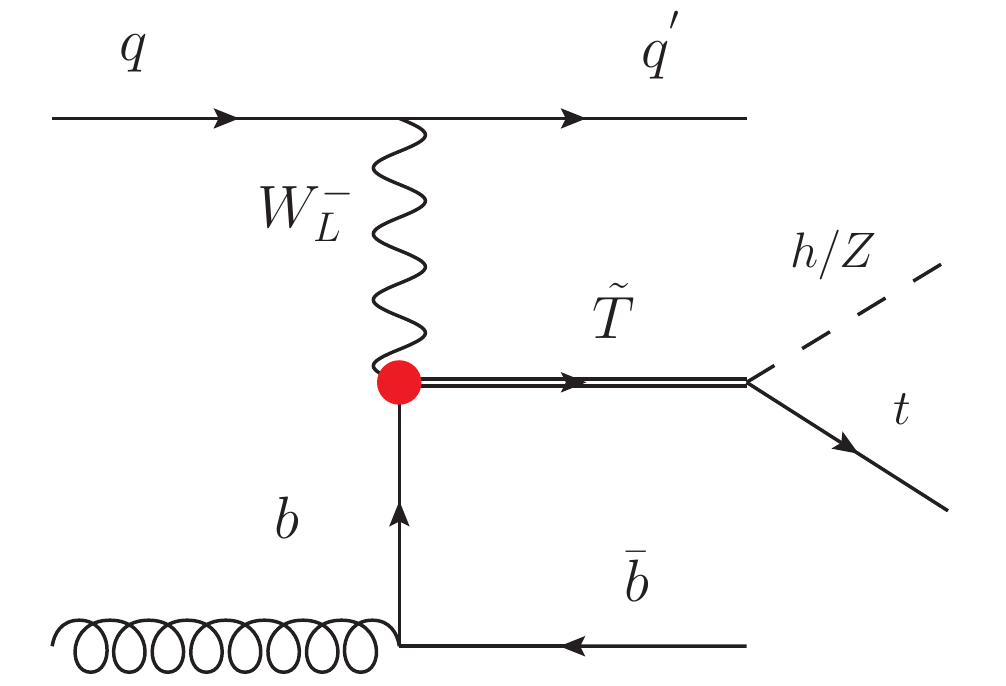}\includegraphics[height=1.3in]{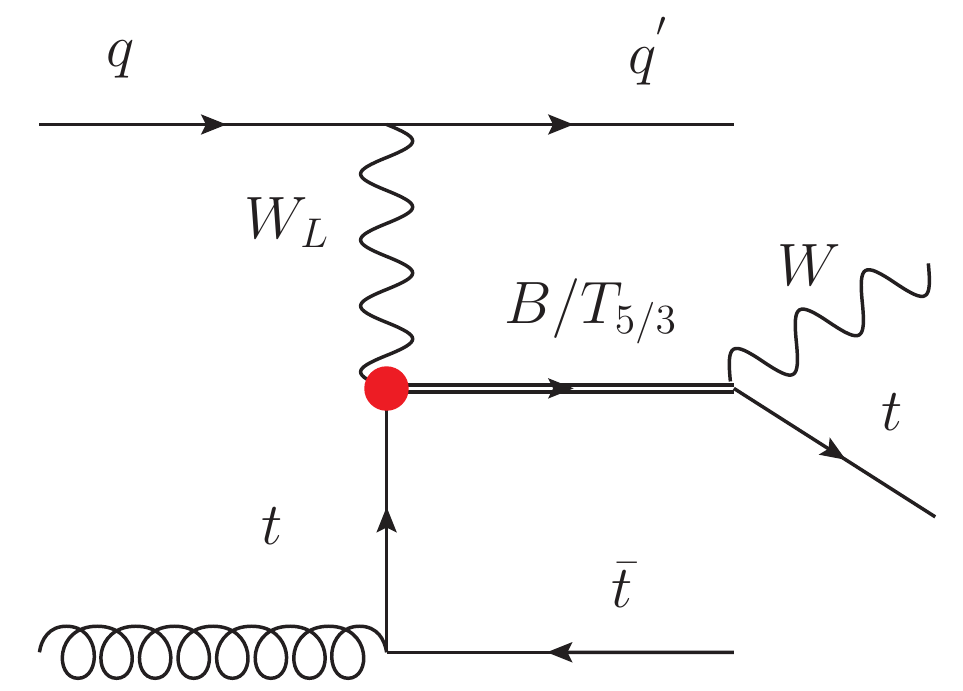}\includegraphics[height=1.1in]{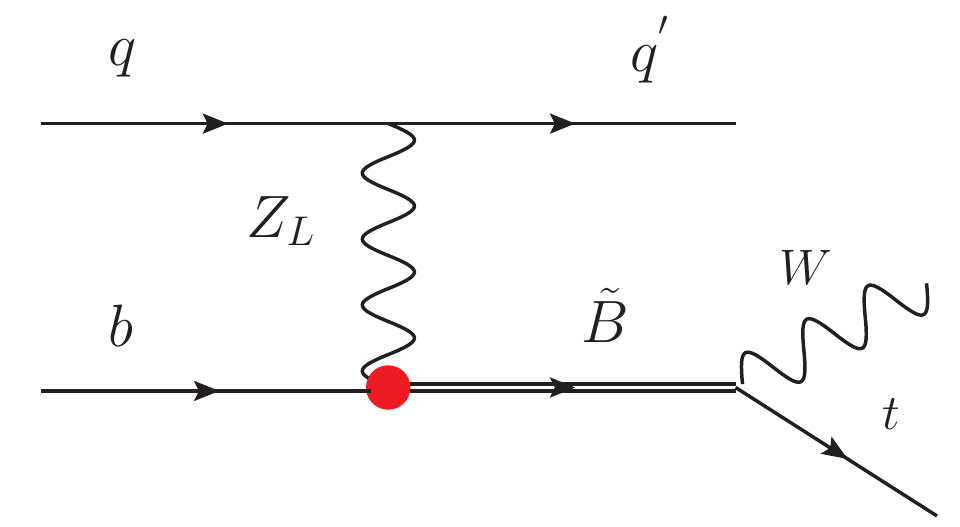}
\caption{\small Single production channels for: (Left) EW singlet $\tilde{T}$, (Middle) $B$ or $T_{5/3}$, (Right) EW singlet bottom-partner (predicted, for example, in models with VLQs in a 10 of $SO(5)$ \cite{Vignaroli:2012si}).}
\label{fig:vlq-single}
\end{figure}

In minimal composite Higgs models it is considered a scenario of {\it partial compositeness} of the SM fermions \cite{Kaplan:1991dc}. Elementary fermions mix linearly with composite VLQs from the new strong sector. The physical states are thus SM fermions and new VLQ heavy fermions which are superpositions of elementary and composite modes. Heavier SM fermions, as the top, have a larger superposition with the strong sector and thus a larger degree of compositeness ($s_L$, $s_R$). \\
The minimal fermionic content of the strong sector, needed to generate the top mass, include an $SU(2)_L$ doublet of $(T\, B)$ VLQs, partner of $q_L=(t_L\, b_L)$, and an EW singlet $\tilde{T}$, partner ot $t_R$. In order to protect the $\rho$ parameter and the $Zb_L \bar{b}_L$ coupling from large corrections, the presence of a custodial symmetry in the strong sector is strongly favored. The minimal fermionic spectrum thus includes an EW bi-doublet of VLQs and we have:
\begin{equation}
\left( \begin{array}{cc} 
T & T_{5/3} \\ B & T_{2/3}
\end{array}\right)\equiv (2,2)_X   \qquad \tilde{T}\equiv (1,1)_X \, \, ,
\end{equation}
where we have indicated the representations under $SU(2)_L\times SU(2)_R \times U(1)_X$ (the hypercharge is realized as $Y=X+T^3_R$). The additional doublet of exotic VLQs, $(T_{5/3} \, T_{2/3})$, thus appears as a consequence of the custodial symmetry. The ``custodian" VLQs, that cannot mix directly with $q_L$, are lighter than $(T\, B)$. The difference in mass increases for larger $t_L$ degree of compositeness. In the limit $\lambda/m_{VLQ}\ll 1$, with $\lambda$ the Yukawa coupling between VLQs and EW Goldstone bosons, we have the pattern of decay BRs: \footnote{Outside the $\lambda/m_{VLQ}\ll 1$ regime a more complete description, including the full diagonalization of the fermionic mass matrices, is needed and deviations from the pattern of BRs predicted by the equivalence theorem may be relevant \cite{Aguilar-Saavedra:2013qpa}. }
 
 \begin{equation}
 \tilde{T} \to Wb , \,  Zt , \,  ht \, \, (2:1:1) \qquad T, T_{2/3} \to Zt, \, ht \, \, (1:1) \qquad B, T_{5/3} \to Wt \, .
 \end{equation}
Tops are thus produced by the decays of the VLQs. VLQs can be produced at the LHC in pairs. The production in this case is purely regulated by the SM QCD group. An alternative production mode is also possible: the single EW production \cite{Vignaroli:2012nf,Vignaroli:2012sf}. Searching in the single production channel brings several advantages. First, the cross section is typically enhanced compared to pair production. Then the signal topology is very peculiar, with the presence, for example, of a forward jet in the final state, which can be used to discriminate the signal from the background. Further, it would allow the measurement of the VLQ couplings to EW bosons, thus permitting to gain information on the sector behind the EWSB. Fig. \ref{fig:vlq-single} shows interesting signals of VLQ single production with tops in the final state either coming from the VLQ or radiated from the initial parton. Experimental analyses have been performed in some of the single production channels at the 8 TeV run-1 of the LHC \cite{Aad:2014efa, Aad:2015voa}.   
\section{Vector resonances}

%\begin{wrapfigure}{r}{0.5\textwidth}
%\includegraphics[width=0.4\textwidth]{BR05}
%\caption{\footnotesize $W^{\prime}$ decay BRs. The mass of the custodian VLQ is fixed to 0.9 TeV and the top degree of compositeness to an intermediate value $s_L=0.5$.}
%\vspace{-15pt}
%\label{fig:BR}
%\end{wrapfigure}

\begin{SCfigure}
\centering
\vspace{-20pt}
\caption{\small $W^{\prime}$ decay BRs as function of $m_{W^{\prime}}$. The mass of the custodian VLQ is fixed to 0.9 TeV and the top degree of compositeness to an intermediate value $s_L=0.5$.}
\includegraphics[height=1.7in]{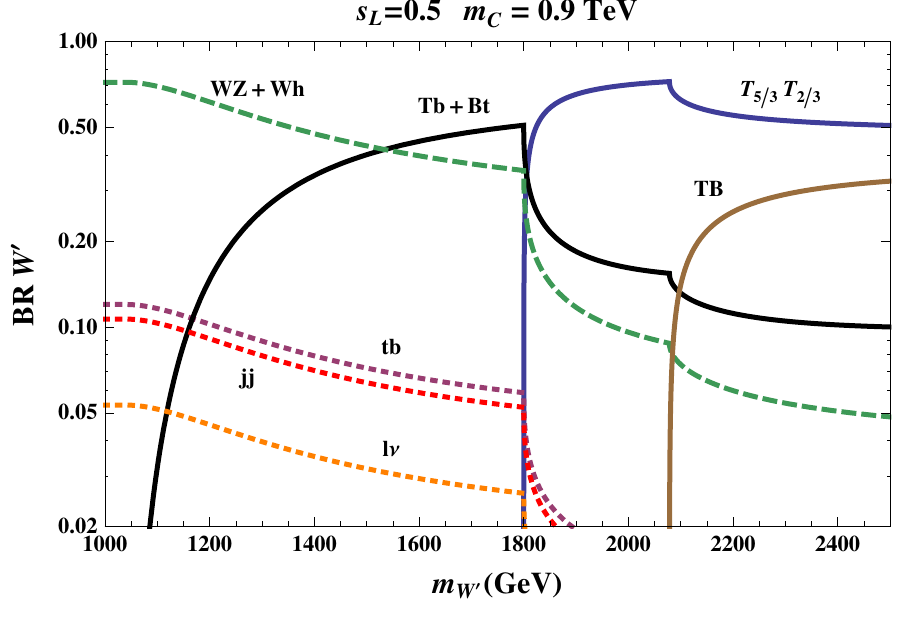}
\label{fig:BR}
\end{SCfigure}

\begin{figure}[htb]
\centering
\begin{subfigure}[b]{0.55\textwidth}
\includegraphics[height=1.1in]{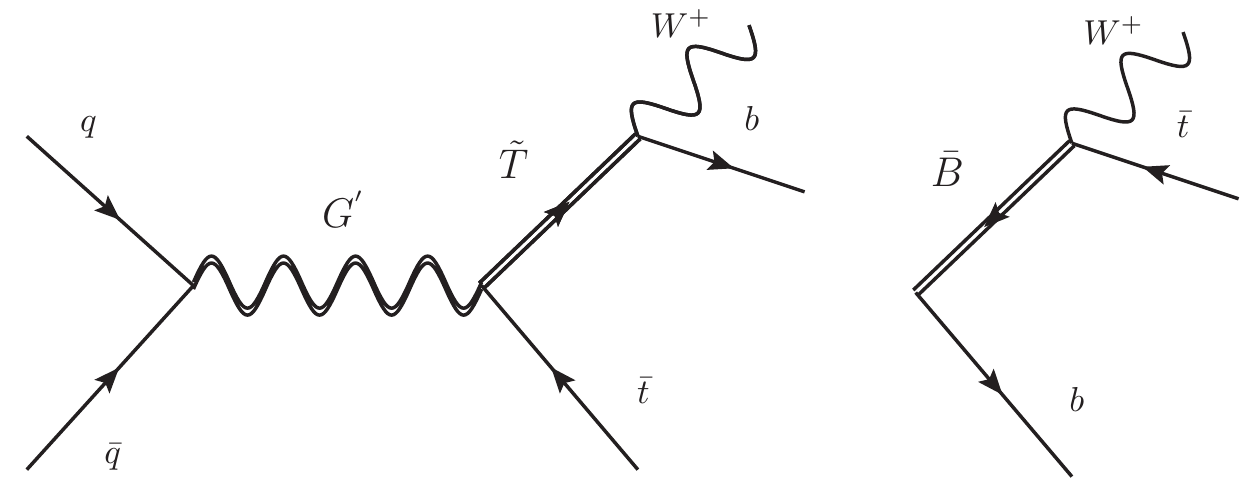}\caption{}
\end{subfigure}
\begin{subfigure}[b]{0.65\textwidth}
\includegraphics[height=1.25 in]{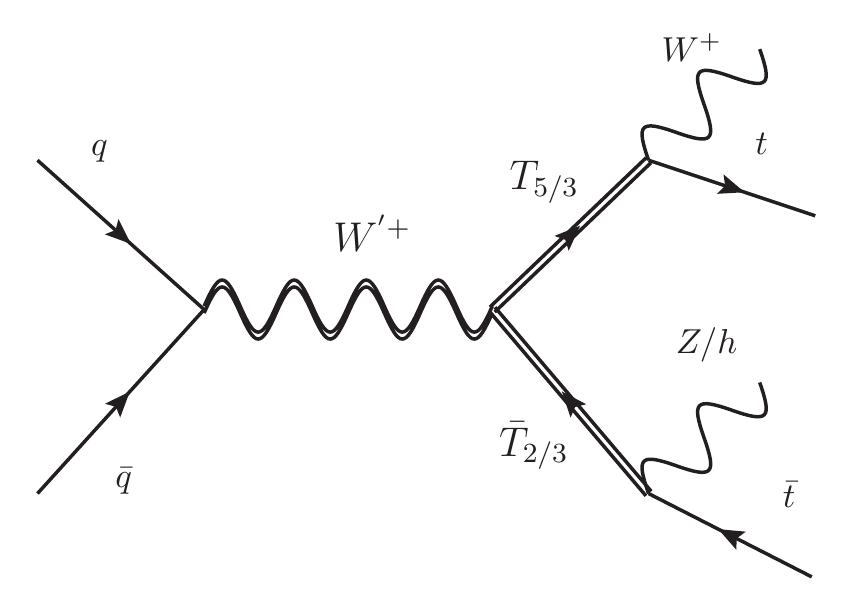}
\includegraphics[height=1.15 in]{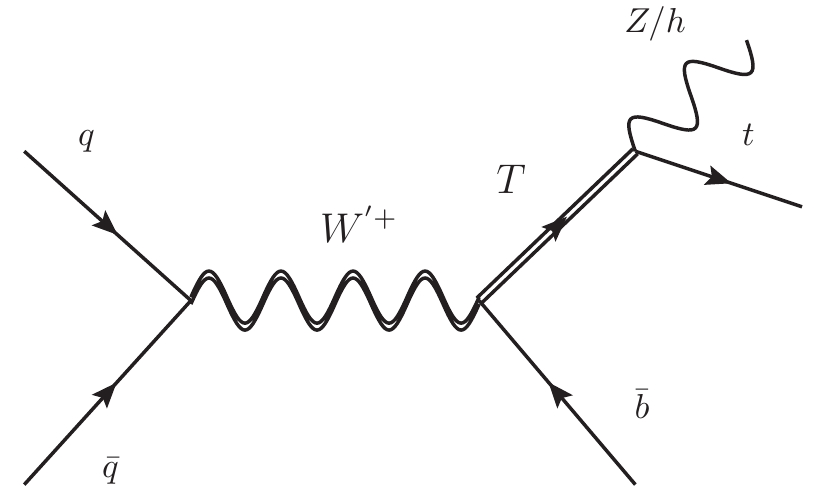}\caption{}
\end{subfigure}
\caption{\small Cascade topologies for vector resonances. (a) heavy-light decay topologies for a heavy gluon in the $Wtb$ final state. Other interesting possible final states include $h/Zt\bar{t}$, $Z/hb\bar{b}$ \cite{Bini:2011zb} or final states coming from $G^{\prime}$ decays into pairs of VLQs \cite{Vignaroli:2015ama}. (b) Cascade topologies for a $W^{\prime}$ \cite{Vignaroli:2014bpa}.}
\label{fig:cascade-top}
\end{figure}

In the partial compositeness framework, below the threshold for decays to VLQs, vector resonances generated from the new strong sector ($W^{\prime}$, $Z^{\prime}$, heavy gluons, etc.) decay dominantly to third-generation quarks and, in the case of EW resonances, to longitudinal $W,Z$ bosons. However, the naturalness argument %(which generally indicates $$) 
as well as indications from electroweak precision data and flavor observables suggest a spectrum where vector resonances are heavier than VLQs \cite{Bini:2011zb, Vignaroli:2014bpa}. Fig. \ref{fig:BR} shows the typical scenario for the decay BRs of a $W^{\prime}$ resonance (similar BRs are found for other types of resonances, $Z^{\prime}$, $G^{\prime}$, etc.), with dominant decays to VLQs. It is thus interesting to search at the LHC for ``cascade" topologies as those in Fig. \ref{fig:cascade-top}, where vector resonances decay to one or pairs of VLQs and the tops are typically generated by the decays of the VLQs. The main production mechanism at the LHC is the Drell-Yan. For the EW resonances, the vector-boson-fusion channel can also become relevant at a futuristic 100 TeV $pp$ collider \cite{Mohan:2015doa, Golling:2016gvc}.

\section{New composite scalars}

New composite scalars may also appear in composite Higgs models. For example, in next-to-minimal cosets as $SU(4)/Sp(4)$ or larger, axtra pNGBs are present which may dominantly decay to $t\bar{t}$ \cite{Galloway:2010bp, Cacciapaglia:2014uja, Bellazzini:2015nxw}. The new strong dynamics can also generate $\eta^{\prime}$-like states with anomalous decays to gauge bosons \cite{Molinaro:2015cwg} and which may be directly coupled to tops. The dominant production mode in this case is the top-mediated gluon fusion and the dominant decay is in $t\bar{t}$ \cite{Molinaro:2016oix}. In scenarios where the QCD group emerges from a breaking $SU(3)_1 \times SU(3)_2 \to SU(3)_{\text{QCD}}$, as in top-coloron models \cite{Chivukula:2013kw}, colored scalars associated to the breaking appear, which can be searched at the LHC in final states with tops. In next-to-minimal flavor violating scenarios as in \cite{Chivukula:2013kw, Chivukula:2013hga}, an interesting channel to consider at the LHC is the one with pair produced color-octet scalars, each decaying to $t\bar{c}$ (or $\bar{t}c$), and leading to same-sign dilepton final states.

%%%%%%%%%%%%%%%%%%%%%%%%%%%%%%%%%%%%%%%%%%%%%%%%%%%%%%%%%%%%%%%%%%%%%%%%%
%%
%%   use this format to include an .eps figure into your paper
%%
%\begin{figure}[htb]
%\centering
%\includegraphics[height=1.5in]{magnet}
%\caption{Plan of the magnet used in the mesmeric studies.}
%\label{fig:magnet}
%\end{figure}
%%%%%%%%%%%%%%%%%%%%%%%%%%%%%%%%%%%%%%%%%%%%%%%%%%%%%%%%%%%%%%%%%%%%%%%%%%%

%%%%%%%%%%%%%%%%%%%%%%%%%%%%%%%%%%%%%%%%%%%%%%%%%%%%%%%%%%%%%%%%%%%%%%%%%
%%
%%   use this format to include a LaTeX table  into your paper
%%
%\begin{table}[t]
%\begin{center}
%\begin{tabular}{l|ccc}  
%Patient &  Initial level($\mu$g/cc) &  w. Magnet &  
%w. Magnet and Sound \\ \hline
% Guglielmo B.  &   0.12     &     0.10      &     0.001  \\
% Ferrando di N. &  0.15     &     0.11      &  $< 0.0005$ \\ \hline
%\end{tabular}
%\caption{Blood cyanide levels for the two patients.}
%\label{tab:blood}
%\end{center}
%\end{table}
%%%%%%%%%%%%%%%%%%%%%%%%%%%%%%%%%%%%%%%%%%%%%%%%%%%%%%%%%%%%%%%%%%%%%%%%%%%

%\Acknowledgements


\begin{thebibliography}{99}

%%
%%  bibliographic items can be constructed using the LaTeX format in SPIRES:
%%    see    http://www.slac.stanford.edu/spires/hep/latex.html
%%  SPIRES will also supply the CITATION line information; please include it.
%%

{\small

%\cite{Kaplan:1983fs}
\bibitem{Kaplan:1983fs} 
  D.~B.~Kaplan and H.~Georgi,
  %``SU(2) x U(1) Breaking by Vacuum Misalignment,''
  Phys.\ Lett.\  {\bf 136B}, 183 (1984).
  doi:10.1016/0370-2693(84)91177-8
  %%CITATION = doi:10.1016/0370-2693(84)91177-8;%%
  %571 citations counted in INSPIRE as of 07 Nov 2016
  
  %\cite{Agashe:2004rs}
\bibitem{Agashe:2004rs} 
  K.~Agashe, R.~Contino and A.~Pomarol,
  %``The Minimal composite Higgs model,''
  Nucl.\ Phys.\ B {\bf 719}, 165 (2005)
  doi:10.1016/j.nuclphysb.2005.04.035
  [hep-ph/0412089].
  %%CITATION = doi:10.1016/j.nuclphysb.2005.04.035;%%
  %883 citations counted in INSPIRE as of 07 Nov 2016

%\cite{Matsedonskyi:2012ym}
\bibitem{Matsedonskyi:2012ym} 
  O.~Matsedonskyi, G.~Panico and A.~Wulzer,
  %``Light Top Partners for a Light Composite Higgs,''
  JHEP {\bf 1301}, 164 (2013)
  doi:10.1007/JHEP01(2013)164
  [arXiv:1204.6333 [hep-ph]].
  %%CITATION = doi:10.1007/JHEP01(2013)164;%%
  %135 citations counted in INSPIRE as of 07 Nov 2016
  
  %\cite{Kaplan:1991dc}
\bibitem{Kaplan:1991dc} 
  D.~B.~Kaplan,
  %``Flavor at SSC energies: A New mechanism for dynamically generated fermion masses,''
  Nucl.\ Phys.\ B {\bf 365}, 259 (1991).
  doi:10.1016/S0550-3213(05)80021-5
  %%CITATION = doi:10.1016/S0550-3213(05)80021-5;%%
  %273 citations counted in INSPIRE as of 10 Nov 2016


%\cite{Aguilar-Saavedra:2013qpa}
\bibitem{Aguilar-Saavedra:2013qpa} 
  J.~A.~Aguilar-Saavedra, R.~Benbrik, S.~Heinemeyer and M.~Pérez-Victoria,
  %``Handbook of vectorlike quarks: Mixing and single production,''
  Phys.\ Rev.\ D {\bf 88}, no. 9, 094010 (2013)
  doi:10.1103/PhysRevD.88.094010
  [arXiv:1306.0572 [hep-ph]].
  %%CITATION = doi:10.1103/PhysRevD.88.094010;%%
  %161 citations counted in INSPIRE as of 07 Nov 2016

%\cite{Vignaroli:2012nf}
\bibitem{Vignaroli:2012nf} 
  N.~Vignaroli,
  %``Early discovery of top partners and test of the Higgs nature,''
  Phys.\ Rev.\ D {\bf 86}, 075017 (2012)
  doi:10.1103/PhysRevD.86.075017
  [arXiv:1207.0830 [hep-ph]].
  %%CITATION = doi:10.1103/PhysRevD.86.075017;%%
  %38 citations counted in INSPIRE as of 07 Nov 2016
  
%\cite{Vignaroli:2012sf}
\bibitem{Vignaroli:2012sf} 
  N.~Vignaroli,
  %``Discovering the composite Higgs through the decay of a heavy fermion,''
  JHEP {\bf 1207}, 158 (2012)
  doi:10.1007/JHEP07(2012)158
  [arXiv:1204.0468 [hep-ph]].
  %%CITATION = doi:10.1007/JHEP07(2012)158;%%
  %38 citations counted in INSPIRE as of 07 Nov 2016
  
%\cite{Vignaroli:2012si}
\bibitem{Vignaroli:2012si} 
  N.~Vignaroli,
  %``$\Delta$ F=1 constraints on composite Higgs models with LR parity,''
  Phys.\ Rev.\ D {\bf 86}, 115011 (2012)
  doi:10.1103/PhysRevD.86.115011
  [arXiv:1204.0478 [hep-ph]].
  %%CITATION = doi:10.1103/PhysRevD.86.115011;%%
  %36 citations counted in INSPIRE as of 09 Nov 2016

%\cite{Aad:2014efa}
\bibitem{Aad:2014efa} 
  G.~Aad {\it et al.} [ATLAS Collaboration],
  %``Search for pair and single production of new heavy quarks that decay to a $Z$ boson and a third-generation quark in $pp$ collisions at $\sqrt{s}=8$ TeV with the ATLAS detector,''
  JHEP {\bf 1411}, 104 (2014)
  doi:10.1007/JHEP11(2014)104
  [arXiv:1409.5500 [hep-ex]].
  %%CITATION = doi:10.1007/JHEP11(2014)104;%%
  %76 citations counted in INSPIRE as of 10 Nov 2016

%\cite{Aad:2015voa}
\bibitem{Aad:2015voa} 
  G.~Aad {\it et al.} [ATLAS Collaboration],
  %``Search for the production of single vector-like and excited quarks in the $Wt$ final state in $pp$ collisions at $\sqrt{s}$ = 8 TeV with the ATLAS detector,''
  JHEP {\bf 1602}, 110 (2016)
  doi:10.1007/JHEP02(2016)110
  [arXiv:1510.02664 [hep-ex]].
  %%CITATION = doi:10.1007/JHEP02(2016)110;%%
  %20 citations counted in INSPIRE as of 10 Nov 2016

%\cite{Bini:2011zb}
\bibitem{Bini:2011zb} 
  C.~Bini, R.~Contino and N.~Vignaroli,
  %``Heavy-light decay topologies as a new strategy to discover a heavy gluon,''
  JHEP {\bf 1201}, 157 (2012)
  doi:10.1007/JHEP01(2012)157
  [arXiv:1110.6058 [hep-ph]].
  %%CITATION = doi:10.1007/JHEP01(2012)157;%%
  %48 citations counted in INSPIRE as of 07 Nov 2016

  %\cite{Vignaroli:2014bpa}
\bibitem{Vignaroli:2014bpa} 
  N.~Vignaroli,
  %``New W′ signals at the LHC,''
  Phys.\ Rev.\ D {\bf 89}, no. 9, 095027 (2014)
  doi:10.1103/PhysRevD.89.095027
  [arXiv:1404.5558 [hep-ph]].
  %%CITATION = doi:10.1103/PhysRevD.89.095027;%%
  %19 citations counted in INSPIRE as of 07 Nov 2016
  
    
  %\cite{Vignaroli:2015ama}
\bibitem{Vignaroli:2015ama} 
  N.~Vignaroli,
  %``$Z$-peaked excess from heavy gluon decays to vectorlike quarks,''
  Phys.\ Rev.\ D {\bf 91}, no. 11, 115009 (2015)
  doi:10.1103/PhysRevD.91.115009
  [arXiv:1504.01768 [hep-ph]].
  %%CITATION = doi:10.1103/PhysRevD.91.115009;%%
  %20 citations counted in INSPIRE as of 07 Nov 2016


  
  %\cite{Mohan:2015doa}
\bibitem{Mohan:2015doa} 
  K.~Mohan and N.~Vignaroli,
  %``Vector resonances in weak-boson-fusion at future pp colliders,''
  JHEP {\bf 1510}, 031 (2015)
  doi:10.1007/JHEP10(2015)031
  [arXiv:1507.03940 [hep-ph]].
  %%CITATION = doi:10.1007/JHEP10(2015)031;%%
  %2 citations counted in INSPIRE as of 07 Nov 2016
  
  %\cite{Golling:2016gvc}
\bibitem{Golling:2016gvc} 
  T.~Golling {\it et al.},
  %``Physics at a 100 TeV pp collider: beyond the Standard Model phenomena,''
  arXiv:1606.00947 [hep-ph].
  %%CITATION = ARXIV:1606.00947;%%
  %14 citations counted in INSPIRE as of 07 Nov 2016

%\cite{Galloway:2010bp}
\bibitem{Galloway:2010bp} 
  J.~Galloway, J.~A.~Evans, M.~A.~Luty and R.~A.~Tacchi,
  %``Minimal Conformal Technicolor and Precision Electroweak Tests,''
  JHEP {\bf 1010}, 086 (2010)
  doi:10.1007/JHEP10(2010)086
  [arXiv:1001.1361 [hep-ph]].
  %%CITATION = doi:10.1007/JHEP10(2010)086;%%
  %88 citations counted in INSPIRE as of 07 Nov 2016

%\cite{Cacciapaglia:2014uja}
\bibitem{Cacciapaglia:2014uja} 
  G.~Cacciapaglia and F.~Sannino,
  %``Fundamental Composite (Goldstone) Higgs Dynamics,''
  JHEP {\bf 1404}, 111 (2014)
  doi:10.1007/JHEP04(2014)111
  [arXiv:1402.0233 [hep-ph]].
  %%CITATION = doi:10.1007/JHEP04(2014)111;%%
  %57 citations counted in INSPIRE as of 07 Nov 2016
  
  %\cite{Bellazzini:2015nxw}
\bibitem{Bellazzini:2015nxw} 
  B.~Bellazzini, R.~Franceschini, F.~Sala and J.~Serra,
  %``Goldstones in Diphotons,''
  JHEP {\bf 1604}, 072 (2016)
  doi:10.1007/JHEP04(2016)072
  [arXiv:1512.05330 [hep-ph]].
  %%CITATION = doi:10.1007/JHEP04(2016)072;%%
  %184 citations counted in INSPIRE as of 07 Nov 2016
  
  
%\cite{Molinaro:2015cwg}
\bibitem{Molinaro:2015cwg} 
  E.~Molinaro, F.~Sannino and N.~Vignaroli,
  %``Minimal Composite Dynamics versus Axion Origin of the Diphoton excess,''
  Mod.\ Phys.\ Lett.\ A {\bf 31}, no. 26, 1650155 (2016)
  doi:10.1142/S0217732316501558
  [arXiv:1512.05334 [hep-ph]].
  %%CITATION = doi:10.1142/S0217732316501558;%%
  %190 citations counted in INSPIRE as of 07 Nov 2016

%\cite{Molinaro:2016oix}
\bibitem{Molinaro:2016oix} 
  E.~Molinaro, F.~Sannino and N.~Vignaroli,
  %``Collider Tests of (Composite) Diphoton Resonances,''
  Nucl.\ Phys.\ B {\bf 911}, 106 (2016)
  doi:10.1016/j.nuclphysb.2016.07.032
  [arXiv:1602.07574 [hep-ph]].
  %%CITATION = doi:10.1016/j.nuclphysb.2016.07.032;%%
  %18 citations counted in INSPIRE as of 07 Nov 2016
  
  %\cite{Chivukula:2013kw}
\bibitem{Chivukula:2013kw} 
  R.~S.~Chivukula, E.~H.~Simmons and N.~Vignaroli,
  %``A Flavorful Top-Coloron Model,''
  Phys.\ Rev.\ D {\bf 87}, no. 7, 075002 (2013)
  doi:10.1103/PhysRevD.87.075002
  [arXiv:1302.1069 [hep-ph]].
  %%CITATION = doi:10.1103/PhysRevD.87.075002;%%
  %23 citations counted in INSPIRE as of 07 Nov 2016

  %\cite{Chivukula:2013hga}
\bibitem{Chivukula:2013hga} 
  R.~S.~Chivukula, E.~H.~Simmons and N.~Vignaroli,
  %``Same-Sign Dileptons from Colored Scalars in the Flavorful Top-Coloron Model,''
  Phys.\ Rev.\ D {\bf 88}, 034006 (2013)
  doi:10.1103/PhysRevD.88.034006
  [arXiv:1306.2248 [hep-ph]].
  %%CITATION = doi:10.1103/PhysRevD.88.034006;%%
  %11 citations counted in INSPIRE as of 07 Nov 2016

}

\end{thebibliography}
\end{document}